\documentclass[showpacs,preprint,amsmath,amssymb,floatfix,aps,11pt]{revtex4}
\usepackage{graphicx}
\usepackage{graphicx,rotating,lscape}

\begin{document}

\def\erf{\mathrm{erf}}

\title{Open charm meson in nuclear matter at finite temperature beyond the zero range approximation}

\author{C. E. Jim\'{e}nez-Tejero$^1$, A. Ramos$^1$, L. Tol\'{o}s$^2$ and I. Vida\~na$^3$}

\affiliation{$^1$Departament d'Estructura i Constituents de la Mat\`eria and Institut de Ci\`{e}ncies del Cosmos, Universitat de Barcelona, Avda. Diagonal 647, E-08028 Barcelona, Spain}

\affiliation{$^2$ Institut de Ci\`encies de l'Espai (IEEC/CSIC), Campus Universitat
Aut\`onoma de Barcelona, Facultat de Ci\`encies, Torre C5, E-08193 Bellaterra, Spain}

\affiliation{$^3$    Centro de F\'isica Computacional. Department of Physics. University of Coimbra, PT-3004-516, Coimbra (Portugal)}

\begin{abstract}
The properties of open charm mesons, $D$, $\bar D$, $D_s$ and $\bar D_s$ in nuclear matter at finite temperature are studied within a self-consistent
coupled-channel approach. The interaction of the low lying pseudoscalar mesons with the ground state baryons in the charm sector is derived from a
$t$-channel vector-exchange model. The in-medium scattering amplitudes are obtained by solving the Lippmann-Schwinger equation at finite temperature including Pauli blocking effects, as well as $D$, $\bar D$, $D_s$
and $\bar D_s$ self-energies taking their mutual influence into account. We find that the in-medium properties of the $D$ meson are affected by the $D_s$-meson self-energy through the intermediate $D_s Y$ loops coupled to $DN$ states. Similarly, dressing the $\bar{D}$ meson in the $\bar{D}Y$ loops has an influence over the properties of the $\bar{D}_s$ meson.
\end{abstract}

\vspace{0.5cm}

\maketitle


\section{Introduction}

Over the past years the properties of charmed hadrons have received a lot of attention in connection with experiments in lepton colliders (CLEO, Belle, BaBar) and hadron facilities (CDF at Fermilab, PHENIX, STAR at RHIC, and the forthcoming PANDA and CBM experiments at FAIR) \cite{Aubert:2003fg,Briere:2006ff,Krokovny:2003zq,Abe:2003jk,Choi:2003ue,Acosta:2003zx,Abazov:2004kp,Aubert:2004ns,Abe:2007jn,Abe:2007sya,Abe:2004zs,Aubert:2007vj,Uehara:2005qd,2880-Artuso:2000xy,Mizuk:2004yu,Jessop:1998wt,Csorna:2000hw,Chistov:2006zj,Aubert:2007dt,Aubert:2006je,2940-Aubert:2006sp,2880-Abe:2006rz}. 
The discovery of new resonances with charm content has sparked the interest of not only many experimental but also theoretical research groups in order to find plausible explanations for the nature of the newly found states.

The study and characterization of resonances has been a very active topic of research in hadrons physics during the last decades. The goal is to establish
whether some resonances have the genuine $q\bar q$ or $qqq$ structure predicted by the quark model \cite{Godfrey:1985xj,Capstick:1986bm}
  or, alternatively, qualify better as hadron molecules generated dynamically. A series of pioneer works \cite{ball61,Wyld:1967zz,Dalitz:1967fp,Logan:1967zz,raja72,sie88}, based on a $t$-channel vector-meson exchange (TVME) force, and more recent approaches in terms of chiral Lagrangians \cite{Kai95,Nie01,Garcia-Recio:2002td,Ramos:2002xh,ORB02,Jido:2003cb,OR98,Meissner:1999vr,OOR,Oller:2000fj,Nieves:2001wt,Lutz:2001yb,Inoue:2001ip,Oller:2005ig,Borasoy:2005ie,Borasoy:2006sr,Hyodo:2008xr,Hyodo:2007np}
have proven to be very promising and successful in describing a wealth of $S$-wave baryonic resonances in the light SU(3) sector within the molecular picture using coupled-channel dynamics with effective hadronic degrees of freedom. In the modern language of chiral Lagrangians those states emerge from the scattering of the $0^-$ octet Goldstone bosons off baryons of the nucleon $1/2^+$ octet. Moreover, $P-$  and $D-$ wave baryonic resonances have been generated based on the SU(3) leading-order chiral Lagrangian by means of meson scattering off baryons of the $3/2^+$ decuplet \cite{Kolomeitsev:2003kt,Sarkar:2004jh,Roca:2006sz,Doring:2006pt}
 and incorporating vector-meson degrees of freedom \cite{lutz3,Garcia-Recio:2005hy,Toki:2007ab,vijande,souravbao,osetramos}. Molecular states of two pseudoscalar mesons and one baryon  \cite{alberto,alberto2,kanchan,Jido:2008zz,KanadaEn'yo:2008wm}
 have also been studied to interpret low-lying $1/2^+$ states.  All these results sustain the so-called hadrogenesis conjecture, formulated by Lutz and Kolomeitsev a few years ago, according to which resonances not belonging to the large $N_c$ ground state of QCD are generated by coupled-channel dynamics \cite{Lutz:2001yb,hadro}. 

Recently, the charm degree of freedom has been incorporated in those unitarized coupled-channel approaches to describe open- and hidden-charm mesons \cite{Kolomeitsev:2003ac,Hofmann:2003je,Gamermann:2006nm,Gamermann:2007fi,Guo:2006fu,Guo:2006rp}. Similar methods have been used in the meson-baryon sector \cite{Lutz:2003jw,lutz5,Hofmann:2005sw,Tolos:2004yg,Mizutani:2006vq}, partially motivated by the parallelism between the $\Lambda(1405)$ and the  $\Lambda_c(2595)$. The meson-baryon interaction in the charm sector is constructed
using the $t$-channel exchange of vector mesons between pseudoscalar mesons and baryons and performing the zero-range approximation while preserving
chiral symmetry for light mesons \cite{Hofmann:2005sw}. The extension to $D$-wave $J=3/2^-$ resonances was first attempted by extending the basis to include the $J=3/2^+$ baryons \cite{Hofmann:2006qx}. In order to be consistent with the spin-flavor heavy-quark symmetry (HQS) \cite{IW89,Ne94,MW00}, the vector mesons were incorporated later within a SU(8)-inspired model \cite{GarciaRecio:2008dp,Gamermann:2010zz}, similar to the SU(6) one developed in the light sector \cite{Garcia-Recio:2005hy,Toki:2007ab}. An alternative approach based on the local hidden gauge formalism has recently become available \cite{Wu:2010jy}.
 On the other hand, there have been also attempts to
construct the $DN$ and $\bar DN$ interaction by incorporating the charm degree of freedom in the SU(3) meson-exchange model of the J\"ulich group \cite{Haidenbauer:2007jq,Haidenbauer:2008ff,Haidenbauer:2010ch}.

Nuclear medium modifications have been lately incorporated as a second step. The aim is to further investigate on the nature of resonant states, such as $\Lambda_c(2595)$, but also to test the dynamics of charmed hadrons with nucleons and nuclei. The properties of open-charm mesons  in nuclear matter can influence the charmonium production in hot dense matter, which might  indicate the formation of the quark-gluon plasma phase of QCD at high density and temperature \cite{matsui}. Another exciting scenario is the possible formation of $D$-mesic nuclei \cite{Tsushima:1998ru,GarciaRecio:2010vt} and of exotic nuclear bound states like $J/\Psi$ in nuclei \cite{Brodsky:1989jd,Luke92, Tsushima:2011fg}.  From the experimental side, the physics program of the CBM experiment as well as part of the PANDA collaboration at FAIR \cite{fair} will be devoted to the properties of open and hidden charm in dense matter. In particular, the physics goal is to extend to the heavy-quark sector the GSI program for in-medium modifications of hadron properties in the light sector, and to provide insight into the charm-nucleus interaction.

Works based on mean-field approaches provided important shifts for the $D$ and $\bar D$ open-charm meson masses \cite{tsushima,sibirtsev,hayashigaki,mishra}, which alters the formation of
charmonium \cite{Andronic:2007zu}. Some of those models have been recently revised
\cite{Hilger:2008jg,Mishra:2008cd,Kumar:2010gb}. A different perspective is offered by models that, working within coupled-channel unitarized schemes, go beyond mean field and provide the spectral features of the charm mesons in
symmetric nuclear matter at zero \cite{Tolos:2004yg,Mizutani:2006vq,Lutz:2005vx} and finite temperature \cite{Tolos:2005ft,Tolos:2007vh}. Lately, this meson-baryon basis has been extended to incorporate HQS. In this way, not only $D$-meson but also $D^*$-meson 
features have been studied \cite{Tolos:2009nn}.

A common feature of the previous models is the use of an interaction kernel in the zero-range approximation ($t\rightarrow 0$). This is justified for diagonal amplitudes close to threshold and for
non-diagonal transition amplitudes where the masses of mesons and of baryons in the initial and final meson-baryon states differ moderately. However, the charm-exchange processes, for which the difference in masses between the external mesons is comparable with the mass of the charmed vector meson being exchanged, point towards the breakdown of the zero-range approximation. Charmed baryon resonances have been studied using the full $t$-dependence of the $t$-channel vector-exchange driving term in Ref.~\cite{JimenezTejero:2009vq}. Compared to the previous TVME local models, where the $t\to0$ limit was implemented, the work of Ref.~\cite{JimenezTejero:2009vq} obtained the same amount of resonances but located in general at somewhat higher energies and having larger widths. Some of these resonances could clearly be identifiable with experimentally seen states, such as $\Lambda_c(2595)$, $\Sigma_c(2800)$, $\Xi_c(2790)$ and $\Xi_c(2980)$.

In the present work, we study the behavior of the 
dynamically-generated baryonic resonaces in hot dense matter, as well as the spectral features of the open charm mesons ($D$, $\bar D$, $D_s$ and $\bar D_s$),
within a self-consistent coupled-channel approach that considers the full $t$-dependent TVME interaction kernel employed in Ref.~\cite{JimenezTejero:2009vq}. We pay a particular attention to the influence that the dressed mesons exert on each other. We find that the simultaneous dressing of the charm mesons ($D,D_s$) in the $C=1$ sector, or the anticharm mesons  ($\bar{D},\bar{D}_s$) in the $C=-1$ one, affects their in-medium properties in a non-negligible way.

The article is organized as follows. In Sec. \ref{sec:Formalism}, we present
the formalism. We first revise the model adopted for the free space amplitudes
and, next, we describe the modifications that incorporate the medium effects.
Our results for the medium modified resonances and for the spectral functions
of the open-charm mesons at various densities and temperatures are shown in Sec.~III. A summary of our conclusions is
presented in Sec. IV.


\section{Formalism}
\label{sec:Formalism}
In this section, we will first review briefly the coupled-channel approach employed in our previous work \cite{JimenezTejero:2009vq},
where we studied open-charm baryon resonances dynamically generated from the free-space interaction of the low-lying pseudoscalar mesons
with the ground-state baryons using a $t$-channel vector-exchange driving force. After that, we will introduce the main sources of medium effects and we will implement them in our coupled-channel formalism.

Since the properties of the $D$, $\bar{D}$, $D_s$ and $\bar D_s$ mesons in a hot and dense environment will be determined, respectively, from the $DN$,  $\bar{D}N$, $D_s N$ and $\bar D_s N$ amplitudes, we list in Table~\ref{tab:channels} the corresponding set of coupled channels in each of the related isospin ($I$), strangeness ($S$) and charm ($C$) sectors.

\begin{table}[htbp]
    \setlength{\tabcolsep}{0.1cm}
\begin{center}
\begin{tabular}{c|ccccc}

\hline
$(I,S,C)$ &\multicolumn{5}{c}{Channels}\\

\hline
$(\frac{1}{2},-1,-1)$ &  &$\bar{D}_sN(2908)$&$\bar{D}\Lambda(2985)$&$\bar{D}\Sigma(3062)$ &  \\

\hline
$(0,0,-1)$ & & &$\bar{D} N(2806)$ & & \\

\hline
$(1,0,-1)$ & & &$\bar{D} N(2806)$ & & \\

\hline
 $(0,0,1)$   &  $\pi \Sigma_c(2591)$ & $D N(2806)$ & $\eta \Lambda_c(2832)$ & $K\Xi_c(2963)$  & $K\Xi_c'(3070)$ \\
&$D_s\Lambda(3085)$ & $\eta'\Lambda_c(3243)$ & $\eta_c\Lambda_c(5265)$ & $\bar D\Xi_{cc}(5307)$ & \\

\hline
$(1,0,1)$  & $\pi \Lambda_c(2424)$ & $\pi \Sigma_c(2591)$ & $D N(2806)$ & $K\Xi_c(2963)$ & $\eta\Sigma_c(2999)$ \\ &$K\Xi_c'(3070)$ & $D_s\Sigma(3162)$ & $\eta'\Sigma_c(3410)$ & $\bar D\Xi_{cc}(5307)$ & $\eta_c\Sigma_c(5432)$ \\
\hline
$(\frac{1}{2},1,1)$   & & $K \Lambda_c(2779)$ & $D_{s}N(2908)$ &$K \Sigma_c(2946)$   & \\

\hline
\end{tabular}
\end{center}
\caption{Coupled-channel meson-baryon states involved in $DN$, $\bar{D}N$, $D_sN$, or $\bar{D}_sN$ interactions. The energy threshold of each state is given in
brackets.}
\label{tab:channels}
\end{table}

\subsection{Free-space coupled-channel approach}

The free-space amplitudes, $T$, which describe the scattering of the pseudoscalar meson fields off the ground-state baryon fields
can be obtained by solving the well-known Lippmann--Schwinger equation, which schematically reads
\begin{equation}
T=V+VJT\ .
\label{eq:ls}
\end{equation}
The loop function $J$ is the product of the meson and baryon single-particle propagators, and the scattering kernel $V$ describes
the interaction between the pseudoscalar mesons and the ground-state baryons. Following the original work of Hofmann and Lutz \cite{Hofmann:2005sw},
we identify a $t$-channel exchange of vector mesons as the driving force for the $S$-wave scattering between pseudoscalar mesons in 16-plet and baryons
in 20-plet representations. The scattering kernel takes the form (see \cite{Hofmann:2005sw} for details)
\begin{equation}
V^{(I,S,C)}_{ij}(p_i,q_i,p_j,q_j)=\frac{g^2}{4}\sum_{V \in [16]} C^{(I,S,C)}_{ij;V}\bar u(p_j)\gamma^\mu
\left(g_{\mu\nu}-\frac{(q_i-q_j)_\mu (q_i-q_j)_\nu}{m^2_V}\right)
\frac{1}{t-m^2_V}(q_i+q_j)^\nu u(p_i) \ ,
\label{eq:sk}
\end{equation}
where the sum runs over all vector mesons of the SU(4) $16$-plet, $(\rho$, $K^*$, $\bar K^*$, $\omega$, $\phi$,
$D^*$, $D_s^*$, $\bar D^*$, $\bar D_s^*$, $J/\Psi)$, $m_V$ is the mass of the exchanged vector meson, $g$ is the universal vector meson coupling constant, $p_i, q_i,
p_j$ and $q_j$ are the four momenta of the incoming and outgoing baryon and meson, and the coefficients $C^{(I,S,C)}_{ij;V}$ denote the strength of the interaction
in the different $(I,S,C)$ sectors, and meson-baryon channels $(i,j)$. The value of $g=6.6$ that reproduces the decay width of the
$\rho$ meson \cite{pdg} has been considered in this work. The $S$-wave projection of the scattering kernel is easily obtained, and in the center-of-mass (c.m.)
frame it takes the analytical form
\begin{equation}
V_{ij,l=0}^{(I,S,C)}(\vec k_i,\vec k_j)=
N\frac{g^2}{8}
\sum_{V \in[16]}
C^{(I,S,C)}_{ij;V} \left[ \frac{2\beta}{b}+\frac{\alpha b-\beta a}{b^{2}}\ln\left(\frac{a+b}{a-b}\right)\right] \ ,
\label{eq:t}
\end{equation}
with $a, b, \alpha$ and $\beta$ being
\begin{eqnarray}
a &=& m^2_i+m^2_j-2\omega_i(|\vec k_i|) \omega_j(|\vec k_j|)-m^2_V  \, \nonumber \\
b &=& 2 |\vec k_i| |\vec k_j| \, \nonumber \\
\alpha &=& \Omega_i(|\vec k_i|)+\Omega_j(|\vec k_j|)
-M_i-M_j-\frac{m^2_j-m^2_i}{m^2_V}(\Omega_j(|\vec k_j|)-\Omega_i(|\vec k_i|)+M_i-M_j) \, \nonumber \\
\beta &=& \frac{|\vec k_i||\vec k_j| }{(E_i(|\vec k_i|)+M_i)(E_j(|\vec k_j|)+M_j)}
\left(\Omega_i(|\vec k_i|)+\Omega_j(|\vec k_j|)+M_i+M_j \phantom{-\frac{m^2_j-m^2_i}{m^2_V}}
\right. \nonumber \\
&& \phantom{\frac{|\vec k_i||\vec k_j|}{(E(|\vec k_i|)+M_i)(E(|\vec k_j|)+M_j)} }
\left. -\frac{m^2_j-m^2_i}{m^2_V}(\Omega_j(|\vec k_j|)-\Omega_i(|\vec k_i|)+M_j-M_i)\right)
\ ,
\end{eqnarray}
where $\vec k_i,\vec k_j$ are the initial and final relative momenta, $m_i,m_j,M_i,M_j$ are the masses
of the incoming and outgoing mesons and baryons, and $\omega_i(|\vec k_i|),\omega_j(|\vec k_j|),
E_i(|\vec k_i|), E_j(|\vec k_j|)$ their corresponding energies, which have been taken to be their
on-shell values. The factor $N=[(E(|\vec k_i|)+M_i)(E(|\vec k_j|)+M_j)/(4M_i M_j)]^{1/2}$
comes from the normalization of the Dirac spinors. We have defined $\Omega(|\vec k|)\equiv\omega(|\vec k|)+ E(|\vec k|)$.
We note that the zero-range approximation ({\it i.e.,} $t\to 0$) of the $S$-wave scattering kernel is obtained by
expanding the logarithm  of Eq.\ (\ref{eq:t}) in the limit $b/a \rightarrow 0$ up to the linear term in $b/a$ and
setting $a=-m^2_V$. The interested reader is referred to our previous work of Ref.~\cite{JimenezTejero:2009vq} for a detailed
analysis of the validity of the zero-range approximation.

In Eqs.\ (\ref{eq:sk}) and (\ref{eq:t}), we have assumed infinitely (zero-width) exchanged vector mesons, because the value of $t$ is never larger than the square of the minimum energy
required for the meson to decay. In
other words, since the mesons being exchanged in this problem are largely off shell, they will be treated as stable particles.

Once the scattering kernel has been constructed, one can finally write the $S$-wave projection  of the  Lippmann--Schwinger equation,
\begin{equation}
T^{(I,S,C)}_{ij,l=0}(\vec k_i,\vec k_j,\sqrt{s})=V^{(I,S,C)}_{ij,l=0}(\vec k_i,\vec k_j)
+\sum_{k}\int\frac{d\vec k}{(2\pi)^3}F(|\vec k|)V^{(I,S,C)}_{ik,l=0}(\vec k_i,\vec k)
J_k(\sqrt{s},\vec k)T^{(I,S,C)}_{kj,l=0}(\vec k,\vec k_j,\sqrt{s}) \,
\label{eq:lspw}
\end{equation}
where $\sqrt{s}$ is the total energy in the c.m. frame. The loop function $J$ explicitly reads
\begin{equation}
J^{(I,S,C)}_k(\sqrt{s},\vec k)=\frac{M_k}{2E_{k}(|\vec k|)\omega_{k}(|\vec k|)}
\frac{1}{\sqrt{s}-E_{k}(|\vec k|)-\omega_{k}(|\vec k|)+i\eta} \ ,
\label{eq:j}
\end{equation}
and  $F(|\vec k|)$ is a dipole-type form factor,
\begin{equation}
F(|\vec k|)=\left(\frac{\Lambda^2}{\Lambda^2+|\vec k|^2}\right)^2 \ ,
\label{eq:ff}
\end{equation}
that has been introduced to regularize the integral. This form is typically adopted in studies of hadron-hadron interactions within the scheme
of Lippmann-Schwinger-type equations in the light flavour sector \cite{machleidt}.  The value of the cut-off $\Lambda$ is a free parameter of our model.
Given the limited amount of data for charmed baryon resonances, and in order to simplify the analysis, the cut-off $\Lambda$ is adjusted to $903$ MeV/c in order to reproduce the
position of the well-known $J^P=1/2^-$ $\Lambda_c(2595)$ having $(I,S,C)=(0,0,1)$, and the same value is used for the other sectors
explored in this work. In Table~\ref{tab:reson}, we summarize the position, width and most important couplings of the dynamically generated states appearing in the various $(I,S,C)$ sectors listed in Table~\ref{tab:channels}. Note that there are no resonances in the singled-channel $(I,S,C)=(0,0,-1)$ and $(1,0,-1)$ sectors of the ${\bar D}N$ interaction. In the other $C=-1$ case, having ($I=\frac{1}{2},S=-1)$, we find a pole just
below the $D_sN$ threshold. The remaining cases have $C=1$ and, although they
were deeply analyzed in Ref.~\cite{JimenezTejero:2009vq}, we briefly comment here a few essential characteristics that will be useful for our discussion of the in-medium results in the next section. In the $(I=0,S=0)$ sector, apart from the $\Lambda_c(2595)$ resonance to which we fit the model, there is another very narrow one at 2805 MeV, just below the threshold for $DN$ states but coupling very little to them. We also predict two narrow resonances in the $(I=1,S=0)$ sector at $2551$ and $2804$ MeV, right below the thresholds of the channels to which they couple more strongly, namely $\pi\Sigma_c$ and $DN$, respectively. In the $(I=\frac{1}{2},S=1)$ case, we predict a cusp-like structure placed at the threshold of $K\Sigma_{c}$, the channel that shows the largest coupling to this state.

\begin{table}[htbp]
    \setlength{\tabcolsep}{0.1cm}
\begin{center}
\begin{tabular}{|c|c|c|c|c|}
\hline $(I,S,C)$ & $M_R$ [MeV] & $\Gamma$ [MeV] & Main decay modes $|g|(channel)$ \\

\hline
$(\frac{1}{2},-1,-1)$&$2906$ (pole) &$0$& $1.3(\overline{D}_sN) $, $1.1(\overline{D}\Lambda) $, $1.9(\overline{D}\Sigma)$ \\

\hline
$(0,0,1)$& $2595$& $0.5$&$0.31(\pi\Sigma_c)^*$, $11(DN)$, $6.0(D_s\Lambda)$,$2.0(\eta_c\Lambda_c)$ \\
         & $2805$&$0.01$&$0.04(\pi\Sigma_c)^*$, $0.27(DN)$, $2.2(\eta\Lambda_c)$, $4.3(K\Xi_c)$, $0.21(D_s\Lambda)$ \\

\hline
$(1,0,1)$&$2551$&$0.16$&$0.05(\pi\Lambda_c)^*$, $3.7(\pi\Sigma_c)$, $1.1(DN)$, $2.1(K\Xi^{'}_c)$ \\
         &$2804$&$5$&$0.27(\pi\Lambda_c)^*$, $0.14(\pi\Sigma_c)^*$, $2.1(DN)$, $1.8(D_s\Sigma)$ \\

\hline
$(\frac{1}{2},1,1)$& 2946 (cusp) & 0.93&$0.002(K\Lambda_c)^*$, $0.03(D_{s}N)^*$, $0.07(K\Sigma_c)$\\

\hline
\end{tabular}
\end{center}
\caption{Dynamically generated baryon resonances with open charm in various $(I,S,C)$ sectors for a cut-off momentum $\Lambda=903$ MeV/c. The table shows the position ($M_R$) and width ($\Gamma$)
of the resonance, together with the couplings to the most important meson-baryon channels, as well as the couplings to the channels in which it can decay (marked with an asterisk). \label{tab:reson}}
\end{table}

\subsection{Medium effects}

The are two main sources of medium effects to consider: one is a consequence of the Pauli exclusion principle, that
prevents the scattering of two nucleons into states which are already occupied. The other is related to the fact that the
properties of all mesons and baryons are modified in the medium due to their interactions with the Fermi sea of nucleons.
Pauli blocking and finite temperature effects can be incorporated in the coupled-channel equations by simply replacing the
free nucleon propagator by the in-medium one,
\begin{equation}
G_N(p_0,\vec p,\rho,T)=\frac{M_N}{E_N(|\vec p\,|)}\left[\frac{1-n_N(\vec p,\rho,T)}{p_0-E_N(|\vec p\,|)+i\epsilon}+\frac{n_N(\vec p,\rho,T)}{p_0-E_N(|\vec p\,|)-i\epsilon}\right]~,
\label{eq:g}
\end{equation}
where $(p_0,\vec p\,)$ is the total four-momentum of the nucleon in the nuclear matter rest frame, $n_N(\vec p,\rho,T)$ is the
usual Fermi--Dirac distribution function, and $E_N(|\vec p\,|)$ is the on-shell energy of the nucleon.

The nuclear medium effects on the mesons can be incorporated by including their corresponding self-energies, $\Pi_m(q_0,\vec q,\rho,T)$,
in the meson propagator
\begin{equation}
D_m(q_0,\vec q,\rho,T)=\frac{1}{q_0^2-\vec{q}\,^2-m_m^2-\Pi_m(q_0,\vec q,\rho,T)}~,
\label{eq:self}
\end{equation}
being $(q_0,\vec q\,)$ the four-momentum of the meson. This is done in practice through the corresponding Lehmann representation
of the meson propagator
\begin{equation}
D_m(q_0,\vec q,\rho,T)=\int^\infty_0 \frac{S_m(\omega,\vec q,\rho,T)}{q_0-\omega+i\epsilon}d\omega
         -\int^\infty_0 \frac{S_{\bar m}(\omega,\vec q,\rho,T)}{q_0+\omega-i\epsilon}d\omega~,
\label{eq:leh}
\end{equation}
where $S_{m(\bar m)}(\omega,\vec q,\rho,T)$ is the spectral function of the meson $m(\bar m)$:
\begin{equation}
S_m(q_0,\vec q,\rho,T)=-\frac{1}{\pi} \mbox{Im}\left( D_m(q_0,\vec q,\rho,T)\right)\\
=-\frac{1}{\pi}\frac{ \mbox{Im}\left( \Pi_m(q_0,\vec q,\rho,T)\right) }{\vert q_0^2-\vec{q}\,^2-m_m^2-\Pi_m(q_0,\vec q,\rho,T)\vert^2}~.
\label{eq:sf}
\end{equation}
We note here that in this work only the $D, \bar D, D_s$ and $\bar D_s$ mesons have been dressed by self-energy insertions. Mesons $\pi, K,\eta,\eta'$
and $\eta_c$ have not been dressed, as done e.g., in Refs.~\cite{Mizutani:2006vq,Tolos:2007vh,Tolos:2009nn}.
The reason is that the states containing these mesons
couple weakly to the $D N$ and $D_s N$ ones and, therefore, it is expected that approximating the $\pi, K,\eta,\eta'$ spectral functions by the free-space ones, {\it i.e.,} delta functions, will not influence much the in-medium properties of the $D$ and $D_s$ mesons. We emphasize, however, that the present work addresses for the first time the simultaneous dressing of the $D$ and $D_s$ mesons in the charm $C=1$ sector, and that of the $\bar{D}$ and $\bar{D}_s$ mesons in the charm $C=-1$ one.

The loop function for the free case given by  Eq.\ (\ref{eq:j}) must now be replaced by the one including the medium and temperature effects on the baryon and meson propagators, as given by Eqs.\ (\ref{eq:g}) and (\ref{eq:leh}). 
Using the Imaginary Time (or Matsubara) Formalism \cite{Matsubara} we obtain:
\begin{eqnarray}
J^{(I,S,C)}_k(P_0,\vec P,\vec k,\rho,T)=\frac{M_k}{E_{k}(|x\vec P+\vec k\,|)}
\left( \int^{\infty}_{0} d\omega~S_m(\omega,y\vec P-\vec k,\rho,T) \frac{1-n(x\vec P+\vec k,\rho,T)+ f(\omega,T)}{P_0-\omega-E_k(|x\vec P+\vec k\,|)+i\epsilon} \right. \nonumber \\
\left.+\int^{\infty}_{0} d\omega~S_{\bar m}(\omega,y\vec P-\vec k,\rho,T) \frac{n(x\vec P+\vec k,\rho,T)+ f(\omega,T)}{P_0+\omega-E_k(|x\vec P+\vec k\,|)-i\epsilon}\right),
\label{eq:j2}
\end{eqnarray}
where $P_0=q_0+E_k(|\vec p\,|), \vec P=\vec q +\vec p$ and $\vec k=y\vec p - x \vec q$, with $x=M_k/(m_k+M_k)$ and $y=m_k/(m_k+M_k)$, are the total energy, total momentum,
and relative momentum of the meson-baryon pair in the nuclear matter rest frame, $n$ is the Fermi distribution of the baryon and $f$ is the Bose enhancement factor of the meson. In practice, given the nuclear densities and temperatures explored in the present work, we can set $f=0$ for all mesons and $n=0$ for all baryons except for nucleons. One might argue that the Bose enhancement factor for the pions should not be ignored. However, as tested in Ref.~\cite{Tolos:2007vh}, the $DN$ amplitudes are insensitive to this factor due to the reduced coupling to $\pi\Sigma_c$ states resulting from the heavy mass of the meson exchanged in the transition potential.

The in-medium scattering amplitudes $T$ are obtained by directly solving the coupled-channel Eq.~(\ref{eq:lspw}) with the medium modified loop
function $J^{(I,S,C)}_m(P_0,\vec P,\vec k,\rho,T)$.  The in-medium self-energies for the $D, \bar D, D_s$ and $\bar D_s$ mesons are then obtained by integrating
the in-medium diagonal scattering amplitudes over the nucleon Fermi sea as
\begin{equation}
\Pi_{D(\bar D)}(q_0,\vec q,\rho,T)=\int \frac{d^3p}{(2\pi)^3} n(\vec p,\rho,T)~[T^{(I=0)}_{D(\bar D)N}(P_0,\vec P,\rho,T)+3T^{(I=1)}_{D(\bar D)N}(P_0,\vec P,\rho,T)]~,
\label{eq:self2}
\end{equation}
for $D$ and $\bar D$, and as
\begin{equation}
\Pi_{D_s(\bar D_s)}(q_0,\vec q,\rho,T)=4 \int \frac{d^3p}{(2\pi)^3} n(\vec p,\rho,T)~T^{(I=1/2)}_{D_s(\bar D_s)N}(P_0,\vec P,\rho,T)~,
\label{eq:self3}
\end{equation}
for $D_s$ and $\bar D_s$.

Finally, we note that the self-energies $\Pi_m$ $(m=D,\bar D,D_s,\bar D_s)$
must be determined in a self-consistent way since they are obtained from the
in-medium scattering amplitudes $T_{DN}$, $T_{\bar D N}$, $T_{D_s N}$ and
$T_{\bar D_s N}$, which contain the loop functions $J^{(I,S,C)}_{DN}$,
$J^{(I,S,C)}_{D_s Y}$ ($DN$ case); $J^{(I,S,C)}_{{\bar D} N}$ (${\bar D N}$
case); $J^{(I,S,C)}_{D_s N}$ (${D_s N}$ case); and $J^{(I,S,C)}_{{\bar D}_s
N}$, $J^{(I,S,C)}_{{\bar D} Y}$ (${\bar D}_s N$ case),  that are themselves
functions of the self-energies  $\Pi_m$.


\section{Results and Discussion}

\begin{figure}[ht!]
\begin{center}
   \includegraphics[width=0.6\textwidth,angle=0]{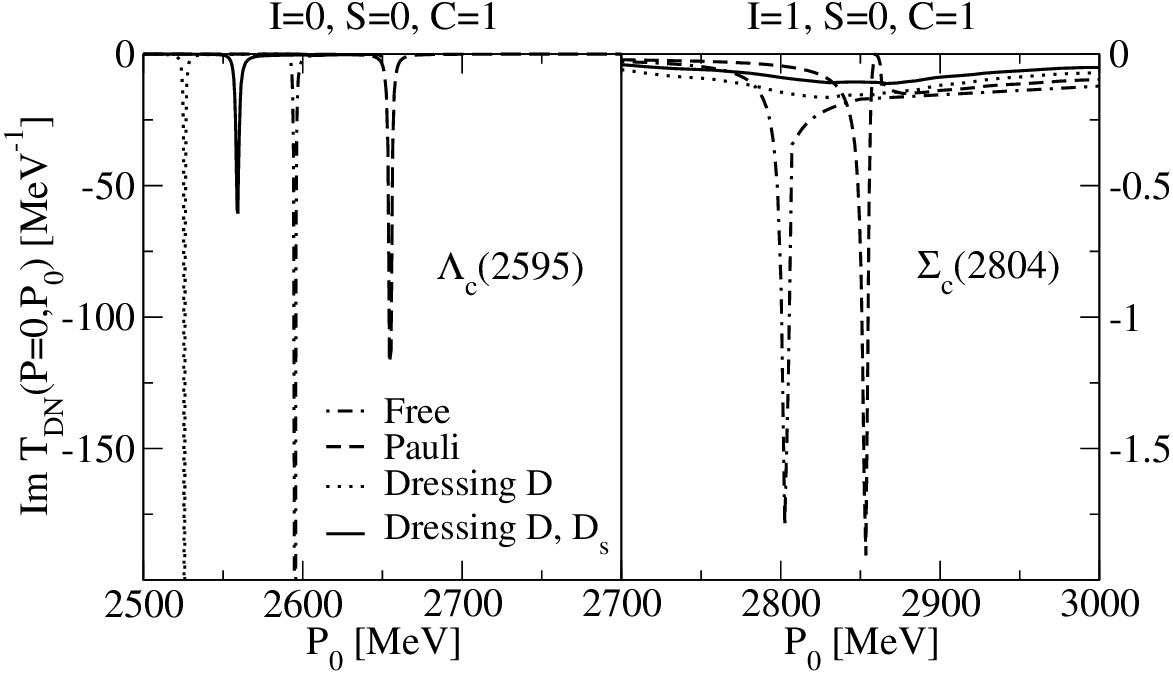}
   \vspace{0.15cm}
   \caption{
   Imaginary part of the  $I=0$ (left panel) and $I=1$ (right panel) $DN\rightarrow DN$ scattering amplitudes in nuclear matter at normal saturation density $\rho_0$ and zero temperature, as function of the total energy $P_0$, for a total momentum $\vec{P}=0$ and various approximations.}
 \label{fig:ampl}
\end{center}
\end{figure}

We will start discussing our results for the $C=1$ mesons, $D$ and $D_s$. First of all, we note that their in-medium properties will be influenced by the charm $C=1$ baryonic resonances that couple significatively to $DN$ and $D_s N$. From the results of our previous work \cite{JimenezTejero:2009vq}, summarized in
Table~\ref{tab:reson}, we find the well known $\Lambda_c(2595)$, coupling very strongly to $DN$ states in the $(I=0,S=0)$ sector, and two other resonances, $\Sigma_c(2551)$ and $\Sigma_c(2804)$, coupling also significatively to $DN$ in the sector $(I=1,S=0)$. The cusp-like structure found in the $(I=1/2,S=1)$ sector shows also a sizable coupling to $D_s N$ states.

In Fig.~\ref{fig:ampl}, we show the imaginary part of the $DN$ amplitude at normal nuclear matter saturation density, $\rho_0=0.17$ fm$^{-1}$, and zero temperature, as a function of the center-of-mass energy $P_0$, covering an energy range that includes the most relevant resonance in each isospin sector, $I=0$ (left panel) and $I=1$ (right panel), for various approximations. The amplitude in free space ($\rho=0$) is also shown (dash-dotted lines), as a reference. When only Pauli blocking effects are included (dashed line) we observe that the $\Lambda_c(2595)$ and  $\Sigma_c(2804)$ states appear displaced to higher energies, by about 60 and 50 MeV, respectively. This repulsive effect is well known, and has to do with the loss of phase space associated to the fact that the nucleons are forced to occupy empty states that are located at momentum states above the Fermi momentum. However, when the dressing of the $D$ meson is incorporated self-consistently (dotted line) $\Lambda_c(2595)$ resonance moves to substantially lower energies and the $\Sigma_c(2804)$ dilutes. This is naturally explained in terms of the $D$-meson strength distribution which, as we will see, shows a quasiparticle peak at a lower energy than in free space and a pronounced peak at even lower energies related to $\Lambda_c(2595)N^{-1}$ excitations. The reduced in-medium $DN$ threshold opens decay channels for the $\Sigma_c(2804)$ which, therefore, broadens considerably. As for the $\Lambda_c(2595)$, its position makes it very sensitive to the low energy strength of the $D$ spectral function and, together with the larger coupling to $DN$ states, explains why the resonance acquires such a large amount of attraction.

In Ref.~\cite{Mizutani:2006vq}, where the TVME in the $t\rightarrow 0$ limit is employed, a similar behavior is observed for the $\Lambda_c(2595)$. The repulsive shift with respect to the free space position due to Pauli blocking effects is compensated by the attractive self-consistent dressing of the $D$ meson. However, the shift is smaller in Ref.~\cite{Mizutani:2006vq}, as it can be seen
from Fig.~5 of this reference (model B). Moreover, this TVME ($t\rightarrow 0$) model also generates a $\Sigma_c$ resonance, which lies around 2795 MeV for model B. This structure melts down as the dressing of $D$ mesons is incorporated, because of the opening of new absortion channels, and stays close to the position with only Pauli blocking effects, in agreement with our present results. In the TVME model of Ref.~\cite{Lutz:2005vx} both resonances are generated but the $\Sigma_c$ one is localized at a much smaller energy, around 2620 MeV. In that work the self-consistent dressing of $D$ mesons results in an attractive shift for both $\Lambda_c$ and $\Sigma_c$ dynamically generated resonances, in contrast to our results.
Within the SU(8) Weinberg-Tomozawa model of Ref.~\cite{GarciaRecio:2008dp}, where heavy-quark spin symmetry is implemented, the $\Sigma_c$ state lies around 2900 MeV and has a different spin, $J=3/2$. In this scheme, the $\Sigma_c$ resonance couples strongly to the $D^*N$ channel instead of $DN$ and behaves similarly  in matter as the $\Lambda_c(2595)$ \cite{Tolos:2009nn}. 

It is clear from Table~\ref{tab:reson} that
the strong coupling of the $\Lambda_c(2595)$ resonance to the
$D_s \Lambda$ channel and that of the $\Sigma_c(2804)$ to $D_s \Sigma$ states, makes it advisable to consider also the medium modifications of the $D_s$ meson in the intermediate $D_s Y$ loops of the
$D N$ amplitude. This is a quite arduous task that, up to our knowledge, has been attempted for the first time in the present work. Our results, represented by the solid lines in Fig.~\ref{fig:ampl}, clearly show the importance of such dressing, making the $\Lambda_c(2595)$ to appear 35 MeV higher in energy with respect to the case of considering free $D_s$ mesons in the $D_s Y$ loops. The changes on the $\Sigma_c(2804)$ resonance are more moderate.

\begin{figure}[ht!]
\begin{center}
 \includegraphics[width=0.6\textwidth,angle=0]{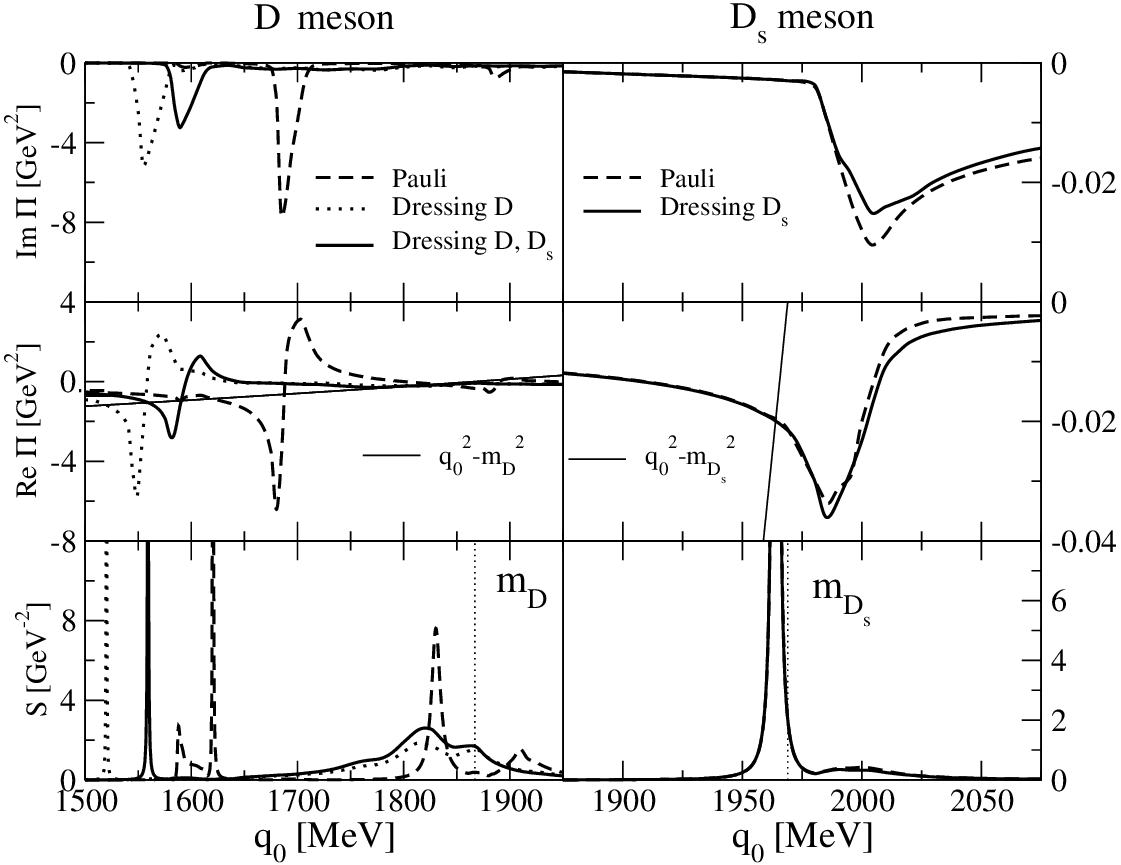}
\caption{Imaginary (upper panels) and real (middle panels) parts of the $D$ (left panel) and $D_s$ (right panel) meson self-energies and
spectral functions (lower panels), as functions of the meson energy $q_0$, at normal nuclear matter saturation density, for $q=0$ MeV/c and
different approximations. The mass of the $D$ meson and the function $q_0^2-m_D^2$ (left panel) and the mass of the $D_s$ meson and the function $q_0^2-m_{D_{s}}^{2}$ (right panel) are shown for comparison.}
\label{fig:selfd}       
\end{center}
\end{figure}

The real and imaginary parts of the $D$ and $D_s$ self-energies and spectral functions at normal nuclear matter saturation density and zero temperature are shown in Fig. \ref{fig:selfd}, as functions of the meson energy, $q_0$, and for
a meson momentum $q=0$ MeV/c. The approximations displayed include: Pauli blocking effects; the additional self-consistent dressing of the given meson; and, in the case of the $D$ meson, the additional dressing of the $D_s$ meson 
in the $D_s Y$ intermediate states coupling to $D N$.

The features discussed for the $DN$ amplitude in Fig.~\ref{fig:ampl} are also reflected in the imaginary part of the $D$-meson self-energy displayed in the upper left panel of Fig.~\ref{fig:selfd}. The middle panel shows the corresponding real part of the self-energy, ${\rm Re}\,\Pi(q_0,\vec{q}=0)$, together with the function $q_0^2-m_D^2$ (thin solid line), such that the crossing points of these two functions signal the appearance of pronounced maxima in the spectral function, as long as the imaginary part does not show a pronounced minimum there.
Actually, the distribution of the $D$-meson strength shown in the lower-left panel is very rich. All the approximations give
a quasi-particle peak located around 35 MeV below the free $D$-meson mass.
In addition, each resonance leaving a signature in the self-energy produces a resonant-hole excitation peak in the spectral function, located at a somewhat different value of energy due to the complex structure of the self-energy. The common behavior is that the resonance-hole modes in the spectral function get displaced such that they move further away from the quasiparticle peak.
In the case of Pauli blocking, we can clearly distinguish three of such modes, associated to ${\Sigma_c}(2551)N^{-1}$,
${\Lambda_c}(2595)N^{-1}$ and ${\Sigma_c}(2804)N^{-1}$ excitations. 
When meson dressing is incorporated, only the ${\Lambda_c}(2595)N^{-1}$ excitation mode is clearly visible. The ${\Sigma_c}(2804)N^{-1}$ mode merges with the quasi-particle peak, and the ${\Sigma_c}(2551)N^{-1}$ one is no longer visible in the spectral function as compared to the ${\Lambda_c}(2595)N^{-1}$ mode. A similar behavior has been observed in Refs.\cite{Mizutani:2006vq,Tolos:2007vh}. In contrast, in Ref.~\cite{Lutz:2005vx}, the 
${\Sigma_c}(2804)N^{-1}$ mode appears at a much lower energy and mixes with the ${\Lambda_c}(2595)N^{-1}$ one, while the quasiparticle peak of the $D$ meson experiences a repulsive shift of 32 MeV. 
It is also  worth mentioning that in the SU(8)-inspired model of Ref.~\cite{Tolos:2009nn} the quasiparticle peak appears at slightly lower energies than the free mass but the $D$-meson spectral function shows a completely different shape due to the different resonant-hole composition of the $D$-meson self-energy.

The imaginary part of the $D_s$ self-energy, displayed in the upper right panel of Fig.~\ref{fig:selfd} shows only a small enhancement at around 2 GeV.
This is a reflection of the enhanced cusp found in the $(I=1/2,S=1,C=1)$ amplitude at the $K\Sigma_c$ threshold \cite{JimenezTejero:2009vq}. This structure generates a small but non-negligible amount of strength in the $D_s$ spectral function to the right of the quasi-particle peak, which barely moves from its free location. This is in contrast to Ref.~\cite{Lutz:2005vx}, where a resonance is generated dynamically 75 MeV below the $D_sN$ threshold, and the corresponding resonance-hole state in the spectral function appears on the left-hand side of the quasiparticle peak.

In spite of the featureless aspect of the $D_s$ spectral function in our model,
this relocation of strength from the quasi-particle peak to higher energies diminishes the size of the $D_s Y$ loops involved in the coupled-channel problem. Therefore, the simultaneous dressing of the $D$  and  $D_s$ mesons in our self-consistent coupled-channel model produces a less bound ${\Lambda_c}(2595)$ resonance in nuclear matter, as already shown in Fig.~\ref{fig:ampl}. From Fig.~\ref{fig:selfd} we can see that the corresponding ${\Lambda_c}(2595)N^{-1}$ excitation mode of the $D$-meson spectral function 
appears approximately 40 MeV higher in energy than when only the $D$-meson dressing is considered.

\begin{figure}[ht!]
\begin{center}
  \includegraphics[width=0.4\textwidth,angle=0]{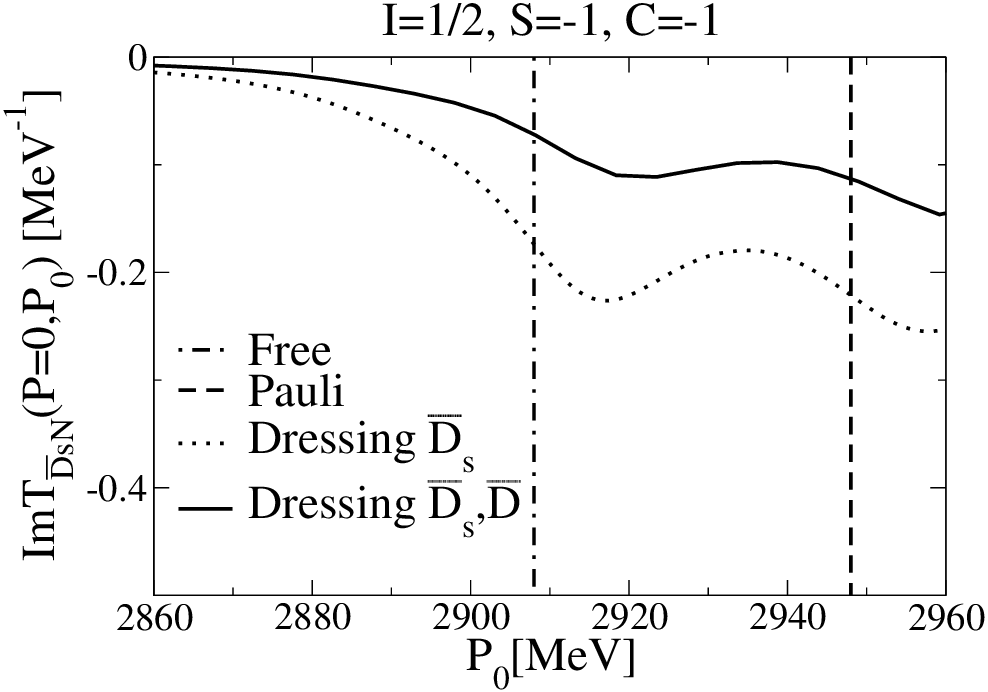}
 \caption{
   Imaginary part of the  $I=1/2$, $S=-1$ and $C=-1$  $\overline{D}_sN \rightarrow \overline{D}_sN$ scattering amplitude in nuclear matter at normal saturation density $\rho_0$ and zero temperature, as a function of the total energy $P_0$, for a total momentum $\vec{P}=0$ and various approximations.}
   \label{fig:ampl2}
\end{center}
\end{figure}

\begin{figure}[ht!]
\begin{center}
  \includegraphics[width=0.6\textwidth,angle=0]{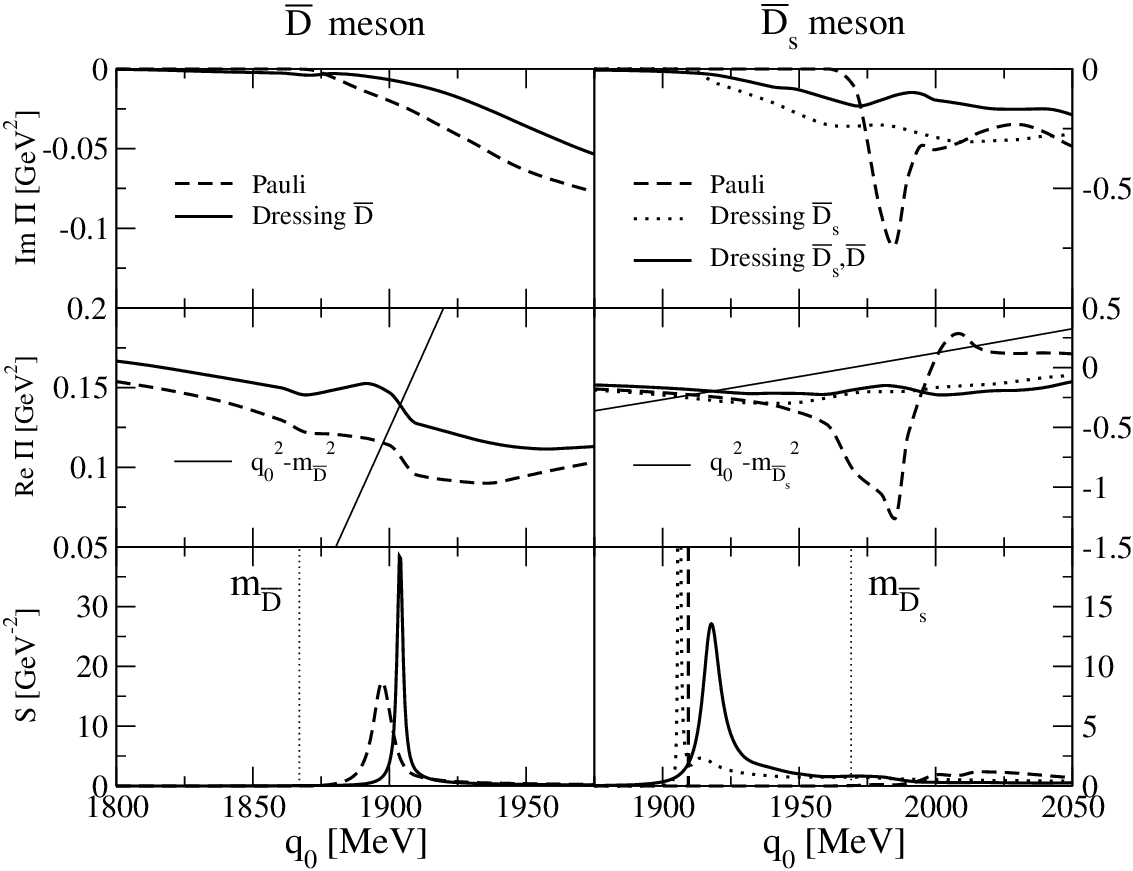}
\caption{The same as Fig.~\ref{fig:selfd} for the
 $\bar D$ (left panels) and $\bar D_s$ (right panels) mesons.}
\label{fig:selfdbar}       
\end{center}
\end{figure}

The in-medium properties of the $C=-1$ mesons, $\bar D$ and $\bar D_s$, will be determined by the behavior of the corresponding $\bar DN$ and $\bar D_sN$ amplitudes in
 the nuclear medium. In Fig.~\ref{fig:ampl2} we display the imaginary part of the $\bar D_sN$ amplitude at normal nuclear matter saturation density and zero temperature as
a function of the center-of-mass energy $P_0$, for various approximations: free (dotted-dashed line), Pauli blocking (dashed line), and the self-consistent calculation including only the dressing of $D_s$ (dotted line) and including both $D_s$ and $D$ dressings (solid line). This is the most interesting of the two $C=-1$ cases since the $\bar D_sN$ system develops in free space a 
subthreshold bound state at $2906$ MeV that couples significatively to $\bar D_sN$ states. Therefore, this pole will be very sensitive to the medium effects. Indeed, when only 
Pauli blocking effects are considered, the pole moves about $40$ MeV towards higher energy  as expected. We observe very drastic changes  when the dressing of the $\bar D$ and $\bar D_s$
mesons is incorporated. The reason is that, as we will see, the in-medium quasiparticle peak of the $\bar D_s$ meson experiences a strong attraction. This moves the 
in-medium threshold for $\bar D_sN$ states below the position of the resonance, making its decay possible and quite probable due to the significant coupling to these states.

The $\bar D$ and $\bar D_s$ self-energies and spectral functions are shown in Fig.~\ref{fig:selfdbar} as functions of $q_0$, including Pauli blocking effects, the additional self-consistent dressing of the given meson, and, in the case of the $\bar D_s$, incorporating also the dressing of the $\bar D$ meson in the $\bar D Y$ intermediate states coupling to $\bar D_s N$. Again the thin solid lines indicate the $q_0^2-m_{\bar D}^{2}$ (left panel) and the $q_0^2-m_{\bar D_s}^{2}$ functions (right panel).

The self-energy of the ${\bar D}$ mesons is quite smooth due to the absence of resonances in the ${\bar D}N$ sector. The repulsive character of the ${\bar D}N$ amplitude gives rise to a quasiparticle peak in the ${\bar D}$ spectral function appearing at higher energy than the ${\bar D}$ meson mass, by 30 MeV in the case of considering Pauli blocking effects only, or by 35 MeV when the additional self-consistent dressing of the ${\bar D}$ meson is also taken into account. The repulsive mass shift obtained in Refs.~\cite{Tolos:2007vh,Lutz:2005vx} is somewhat smaller, of the order of 20 MeV for both cases. On the other hand, the self-energy of the $\bar D_s$ meson shows a richer structure, which, in the case of Pauli-blocking, it is seen as a bump in the spectral function around 2000 MeV.  This is a reflection of the presence, in the ${\bar D_s}N$ amplitude, of a pole at 2906 MeV,  which appears shifted about 40 MeV to higher energies when Pauli blocking effects are incorporated (Fig.~\ref{fig:ampl2}). The dressing of the $\bar D_s$ meson smears this structure in such a way that one barely sees any trace of it in the corresponding spectral function. Moreover, the delta-like quasi-particle peak, appearing 60 MeV below the free ${\bar D_s}$ mass when only Pauli blocking effects are considered, moves to slightly lower energies when the ${\bar D_s}$ meson is dressed. Considering the additional dressing of the $\bar D$ meson in the related ${\bar D}Y$ loops produces a substantial change in the ${\bar D_s}$ self-energy. This is easy to understand from the results of Table~\ref{tab:reson}, where we see that the pole at $2906$ MeV couples also very strongly to $\bar{D}Y$ states. The loss of attraction in the region of the quasiparticle peak moves it towards a higher energy, and ends up being 50 MeV below the free mass and merging with the resonant-hole strength. Our findings differ again quite strongly from those of  Ref.~\cite{Lutz:2005vx}, which are dominated by an exotic coupled-channel molecule at 2780 MeV \cite{Hofmann:2005sw}, which is the equivalent to the pole at $2906$ MeV found in the model of Ref.~\cite{JimenezTejero:2009vq} and used in the present work. 
As a consequence, the spectral function for the ${\bar D}_s$ meson found in Ref.~\cite{Lutz:2005vx} shows two distinct peaks, the quasi-particle one located about 10 MeV above the free $\bar{D}_s$ mass, and a narrow resonance-hole mode located 150 MeV below.

\begin{figure}[ht!]
\begin{center}
 \includegraphics[width=0.6\textwidth,angle=0]{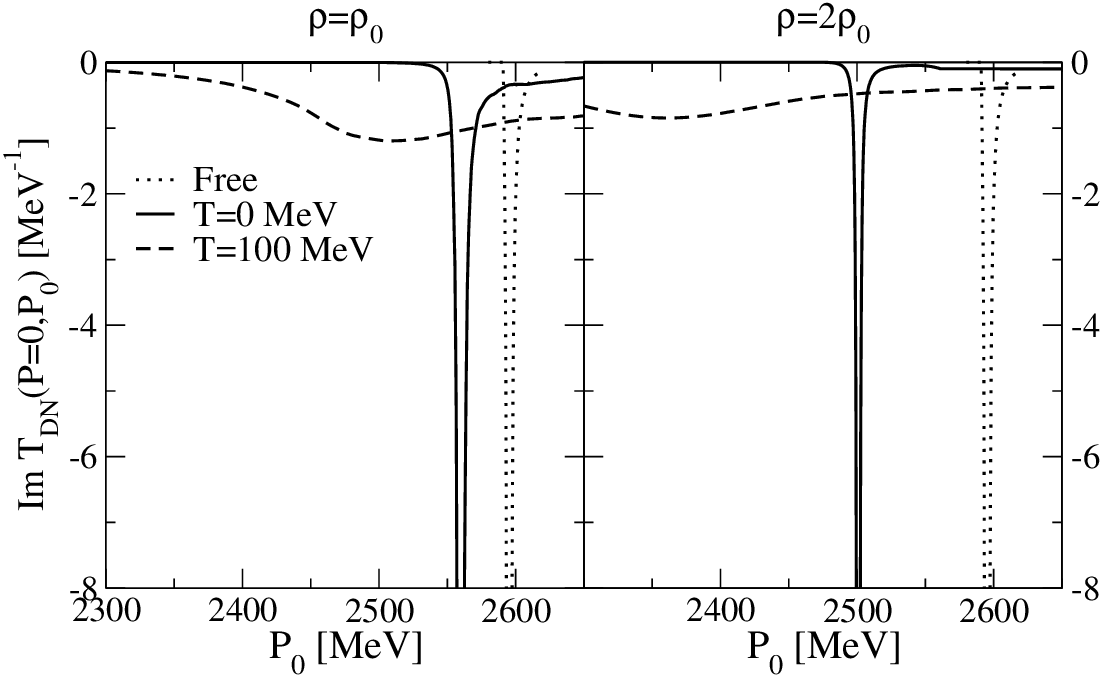}
\caption{Imaginary part of the $I=0$, $DN\rightarrow DN$ scattering amplitudes in nuclear matter at $\rho_0$ (left panel) and $2\rho_0$ (right panel), as functions of the total energy $P_0$, for a total momentum $\vec{P}=0$ and two temperatures, $T=0$ (solid lines) and $T=100$ MeV (dashed lines). The calculation in free space is also given for reference.}
\label{fig:amplt}       
\end{center}
\end{figure}


\begin{figure}[ht!]
\begin{center}
  \includegraphics[width=0.6\textwidth,angle=0]{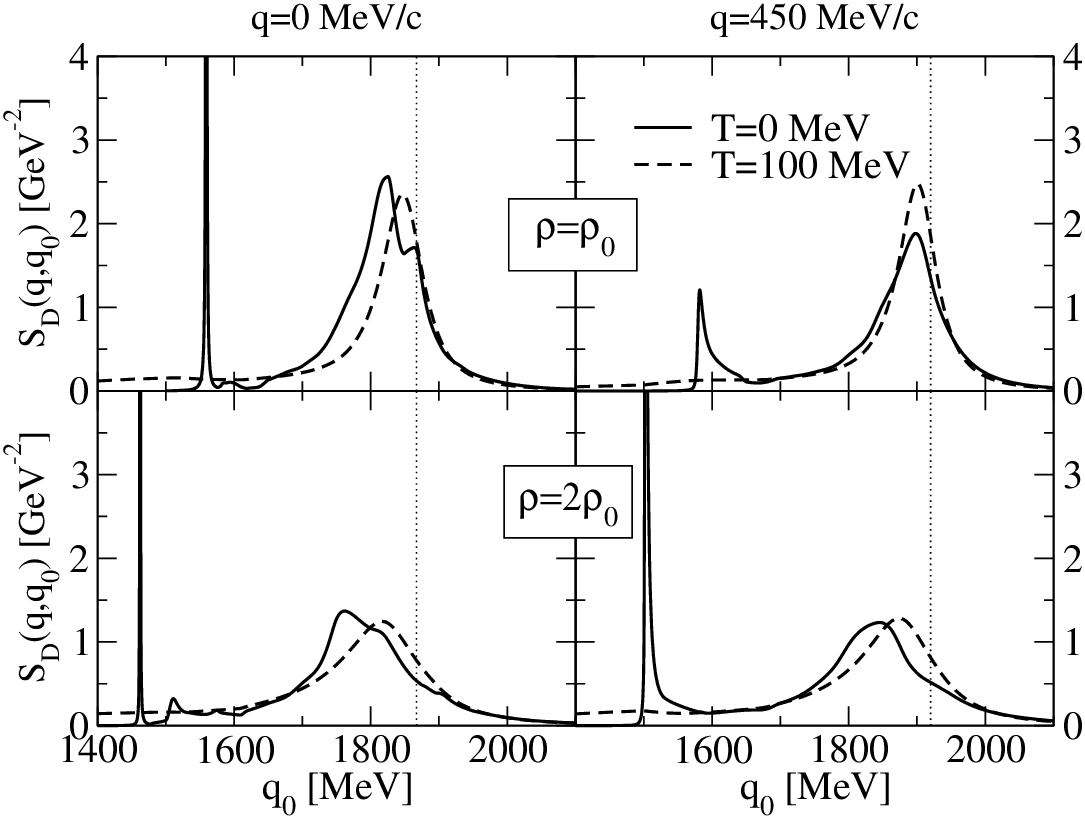}
\caption{The spectral function of the $D$ meson at $\rho=\rho_0$ (upper panels) and $\rho=2\rho_0$ (lower panels), two temperatures, $T=0$ (solid lines) and $T=100$ (dashed lines) and two values of momentum, $q=0$ MeV/c (left panels) and $q=450$ MeV/c (right panels). Dotted lines indicates the free mass of the meson}
\label{fig:specd}       
\end{center}
\end{figure}

\begin{figure}[ht!]
\begin{center}
  \includegraphics[width=0.6\textwidth,angle=0]{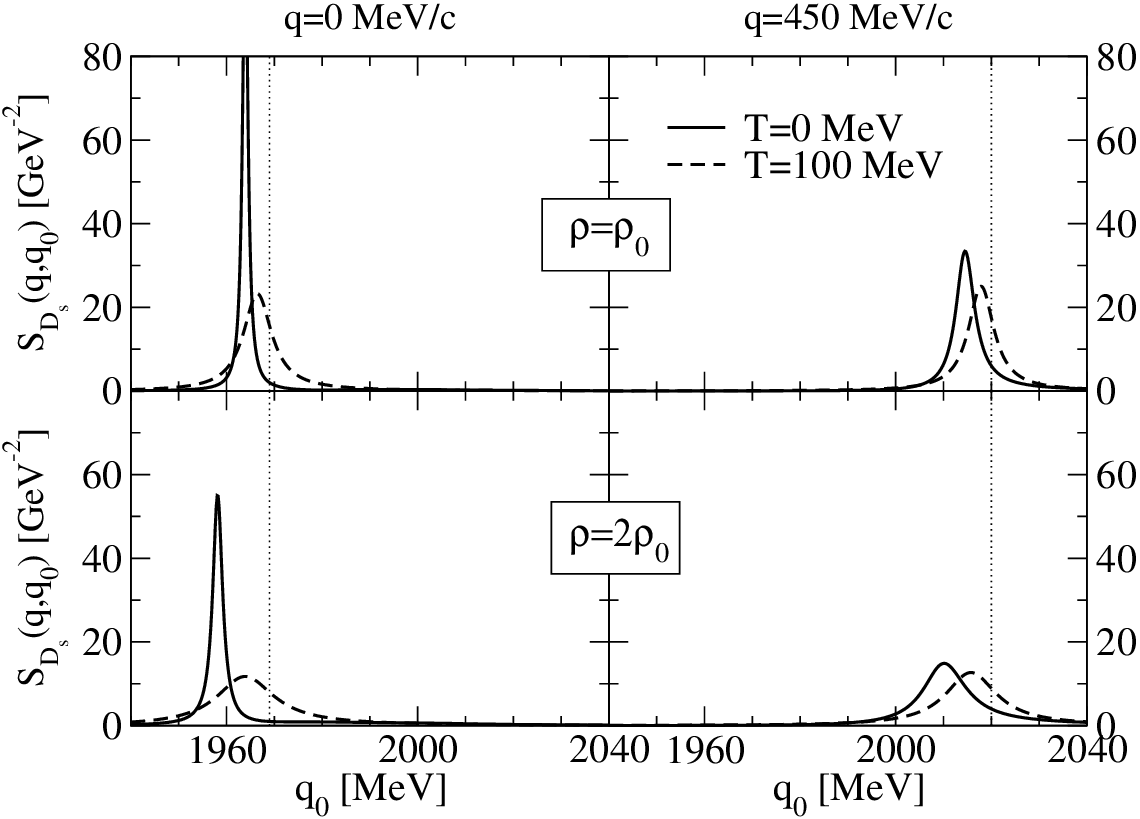}
\caption{The same as Fig.~\ref{fig:specd} for the $D_s$ meson}
\label{fig:specds}       
\end{center}
\end{figure}

\begin{figure}[ht!]
\begin{center}
  \includegraphics[width=0.6\textwidth,angle=0]{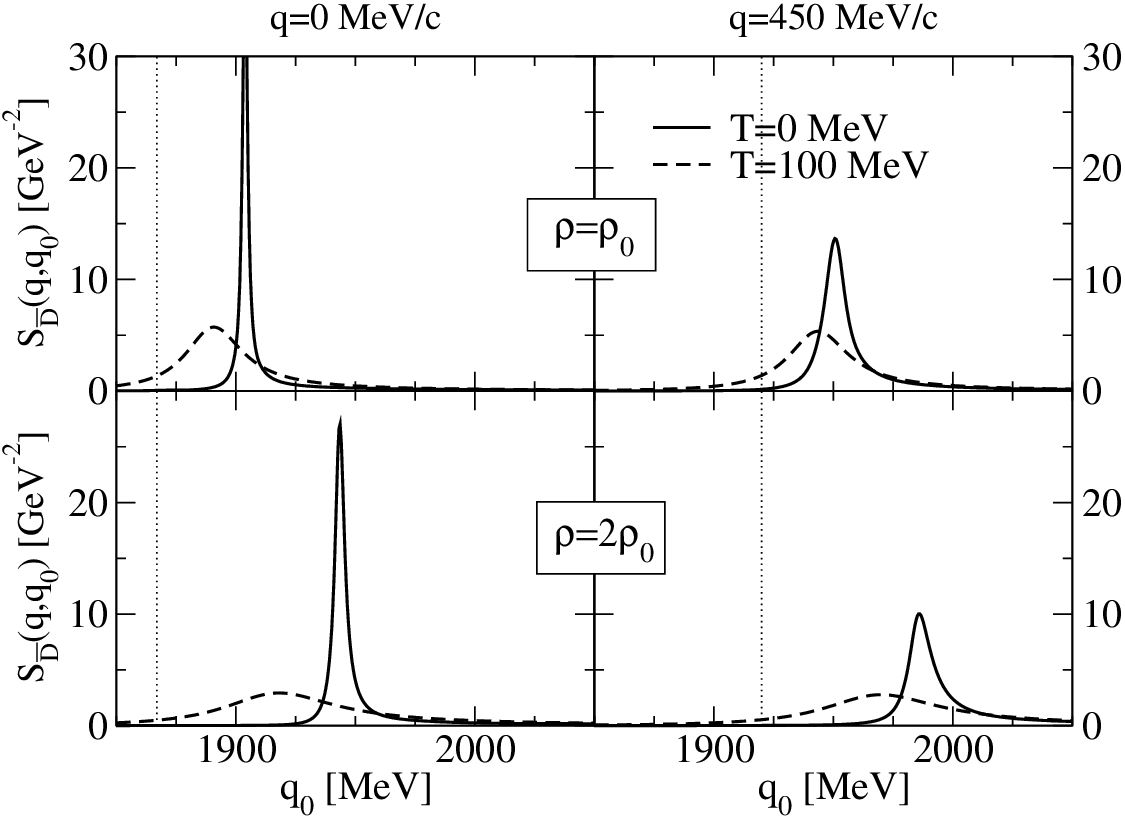}
\caption{The same as Fig.~\ref{fig:specd} for the  $\bar D$ meson.}
\label{fig:specdbar}       
\end{center}
\end{figure}

\begin{figure}[ht!]
\begin{center}
  \includegraphics[width=0.6\textwidth,angle=0]{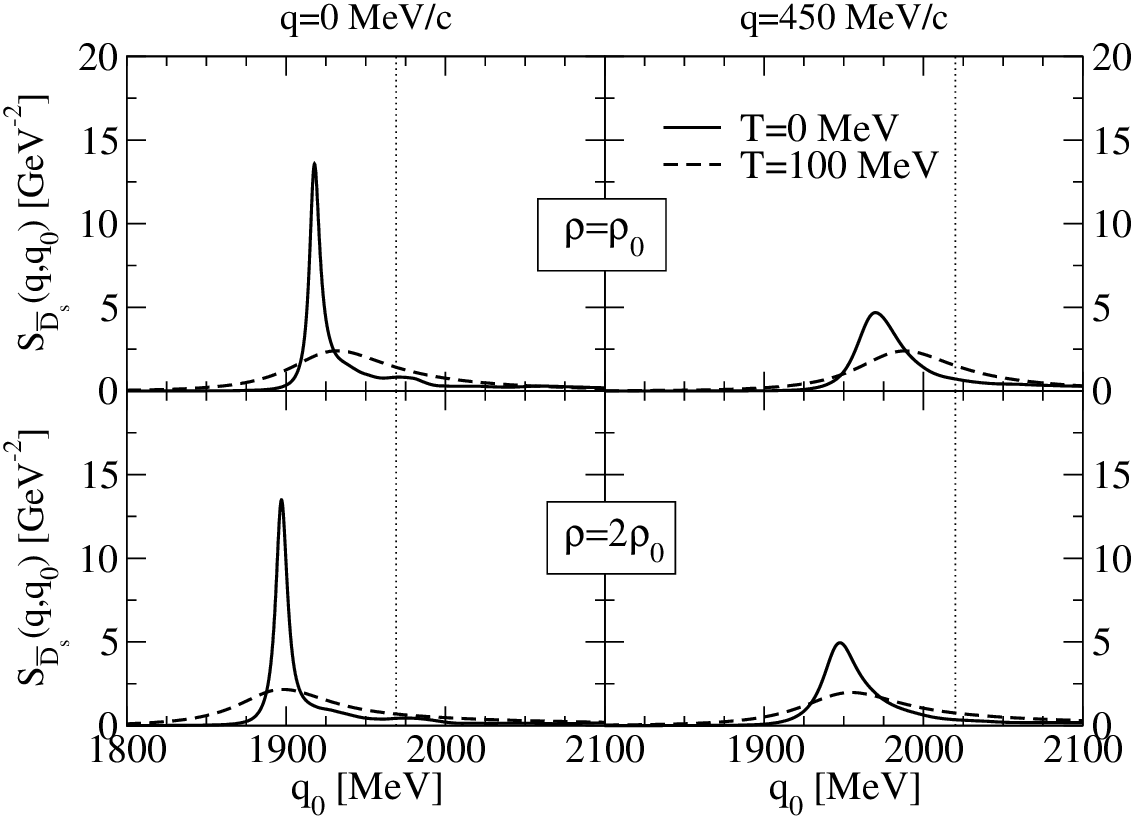}
\caption{The same as Fig.~\ref{fig:specd} for the $\bar D_s$ meson .}
\label{fig:specdbars}       
\end{center}
\end{figure}
In Fig.~\ref{fig:amplt} we display the behavior of the $\Lambda_c(2595)$ resonance at two densities, $\rho_0$ and $2\rho_0$, and two temperatures, $T=0$ and $T=100$ MeV. We observe that the $35$ MeV attraction felt by the resonance at $\rho=\rho_0$ and $T=0$ MeV, gets substantially increased to almost $100$ MeV at twice $\rho_0\rightarrow 2\rho_0$. The changes associated with temperature are also very significant. As already seen in Ref.~\cite{Tolos:2007vh}, increasing the temperature broadens the $\Lambda_c(2595)$ considerably due to the spreading of the $D$-meson strength because of the effect of Fermi motion.

The effect of density and temperature in the spectral functions of the $D$, $D_s$, $\bar D$ and $\bar D_s$ mesons are shown in Figs.~\ref{fig:specd} to \ref{fig:specdbars}, where we compare results for two temperatures, $T=0$ (solid lines) and $T=100$ (dashed lines), and two densities, $\rho=\rho_0$ (upper panels) and $\rho=2\rho_0$ (lower panels), for two values of momentum, $q=0$ MeV/c (left panels) and $q=450$ MeV/c (right panels), in the case of the complete self-consistent calculation, including Pauli blocking and dressing of mesons. A common behavior in all spectral functions is that finite temperature moves the quasiparticle peak towards its free location. This is a reflection of the reduced size of the self-energy, because, being built up from an average over the smeared thermal Fermi distribution, involves higher momentum components for which the meson-nucleon interaction is weaker. Except for a few cases, increasing the temperature gives rise to wider quasiparticle peaks because of the increase of collisional width. However, the opposite effect is seen  for the $D$ meson in Fig.~\ref{fig:specd}. As already discussed in
Ref.~\cite{Tolos:2007vh}, this is due to the fact that the strength under this peak also receives contributions from $\Sigma_c(2804)N^{-1}$ hole excitations, which are washed out by temperature as any other resonant-hole mode. Consequently, the peak of the $D$-meson spectral function becomes narrower and more symmetric as temperature increases, similarly to Ref.~\cite{Tolos:2007vh}.

The density effects observed in the spectral functions are also clearly understood. In general, we find that the self-energy roughly doubles its size when going from nuclear matter at normal nuclear matter saturation density to a system which is two times denser. This is consistent with the low density limit behavior and points at a weak density dependence of the in-medium meson-nucleon amplitude in this density region. This is the reason why, in general, the quasiparticle peak of the spectral functions at $2\rho_0$ are found approximately twice further away from the free space position and are twice wider than in the case of $\rho_0$.



\section{Summary and Conclusions}

We have studied the properties of open charm mesons, $D$, $\bar D$, $D_s$ and $\bar D_s$, in nuclear matter at finite temperature within a self-consistent
coupled-channel approach which uses, as meson-baryon interaction, a full $t$-dependent vector meson exchange driving force.

The in-medium scattering amplitudes are obtained by solving the Lippmann-Schwinger equation at finite temperature including Pauli blocking effects, as well as $D$, $\bar D$, $D_s$
and $\bar D_s$ self-energies, paying a particular attention to their mutual influence. 

We have analyzed how our dynamically generated resonances are affected by density and temperature. As in other similar approaches, the resonances that couple strongly to intermediate states involving nucleons, move upwards in energy when Pauli blocking effects are considered, as a consequence of the loss of phase space. When the self-consistent dressing of the charm mesons is incorporated, the resonances gain attraction again. 

We have seen that dressing the $D_s$ meson has a non-negligible effect on the $D N$ amplitude and on the properties of the $D$ meson. Therefore, we conclude that a simultaneous in-medium treatment of both mesons, as the one attempted in the present work, is necessary. Similarly, the in-medium properties of the $\bar{D}_s$ and $\bar{D}$ mesons are interrelated and must be also considered together.

The spectral functions of the $D$ and $\bar{D}_s$ mesons are quite rich.
 At $T=0$ MeV and normal nuclear matter density one finds a quasiparticle peak located below the corresponding free meson mass by about 50 MeV, as well as strength associated to resonant-hole excitations which, in the particular case of the $D$ meson, is clearly visible as a narrow $\Lambda_c(2595) N^{-1}$ excitation peak.
 
In general, increasing the temperature has the effect of moving the
quasiparticle peak towards its free location making it wider, as a consequence
of a milder meson-baryon interaction and a larger amount of collisions. The exception found for the $D$-meson is naturally explained in terms of the
mixing of the quasi-particle peak with a resonant-hole mode.

For the densities explored, up to twice nuclear matter normal saturation density,  we have found that the density effects follow the linear behavior expected for the low density regime: the self-energy roughly doubles its size when going from nuclear matter at normal saturation density to a system which is two times denser, indicating a mild density dependence of the in-medium meson-baryon interaction amplitudes.

The enormous computational effort of the present work, which uses a
coupled-channel formalism, an interaction that goes beyond the $t\to 0$
limit, and the simultaneous consideration of the in-medium $D$ and $D_s$ 
(${\bar D}_s$ and ${\bar D}$) meson self-energies, has prevented us from
incorporating the coupling to states involving
vector-mesons. We are aware that, given the availability of models that permit
dealing with these important degrees of freedom, 
 our approach should be
extended to the vector mesons such that it also includes, for instance, the
$D^* N$ and $D^*_s Y$ channels in the $C=1$, $S=0$ sector. We hope that, by first
identifying which channels play a relevant role and which ones might be
omitted, we can make progress toward this goal in the nearby future.



\section*{Acknowledgements}
This work is partly supported by the EU contract No. MRTN-CT-2006-035482
(FLAVIAnet), by the contract FIS2008-01661 from MIC
(Spain), by the Ge\-ne\-ra\-li\-tat de Catalunya contract 2009SGR-1289,
and by FEDER/FCT (Portugal) under projects PTDC/FIS/113292/2009 and CERN/FP/109316/2009. We
acknowledge the support of the European Community-Research Infrastructure
Integrating Activity ``Study of Strongly Interacting Matter'' (HadronPhysics2,
Grant Agreement n. 227431) under the Seventh Framework Programme of EU.
L.T. wishes to acknowledge support from the Rosalind Franklin Programme
of the University of Groningen (The Netherlands) and the Helmholtz
International Center for FAIR within the framework of the LOEWE
program by the State of Hesse (Germany). This work has been partially completed thanks to the HPC cluster Millipede of the University of Groningen.


\end{document}